\documentclass[aps, prl, 10pt, twocolumn, superscriptaddress,noshowpacs, preprintnumbers, longbibliography, bibnotes,hyperref,floatfix]{revtex4-1}

\usepackage[colorlinks=true,breaklinks=true]{hyperref}
\hypersetup{allcolors=[rgb]{0.0 0.0 1.0},linkcolor=[rgb]{0.75 0.05 0.05}}
\usepackage[dvipsnames]{xcolor}
\usepackage{mathtools}
\usepackage{amsmath}
\usepackage{amsfonts}
\usepackage{amssymb}
\usepackage{bbold}
\usepackage{bm}
\usepackage{hyperref}
\long\def\exclude#1{}
\usepackage{orcidlink}

\newcommand{\bF}{{\bf F}}

\newcommand{\bp}{{\bf p}}
\newcommand{\br}{{\bf r}}
\newcommand{\bk}{{\bf k}}
\newcommand{\bq}{{\bf q}}
\newcommand{\bx}{{\bf x}}

\newcommand{\ba}{{\bf a}}
\newcommand{\bkk}{{\bf k}}
\newcommand{\bPhi}{{\boldsymbol{\Phi}}}
\newcommand{\bb}{{\bf b}}

\newcommand{\be}{{\bf e}}

\newcommand{\vP}{{\vec{P}}}

\newcommand{\bv}{{\bf v}}

\newcommand{\GF}{G_{\rm F}}

\begin{document}

\title{Inhomogeneous Kinetic Equation for Mixed Neutrinos: Tracing the Missing Energy}

\author{Damiano F.\ G.\ Fiorillo \orcidlink{0000-0003-4927-9850}} 
\affiliation{Niels Bohr International Academy, Niels Bohr Institute,
University of Copenhagen, 2100 Copenhagen, Denmark}

\author{Georg G.\ Raffelt
\orcidlink{0000-0002-0199-9560}}
\affiliation{Max-Planck-Institut f\"ur Physik (Werner-Heisenberg-Institut),  Boltzmannstr.~8, 85748 Garching, Germany}

\author{G\"unter Sigl
\orcidlink{0000-0002-4396-645X}}
\affiliation{Universit\"at Hamburg, II.~Institut f\"ur Theoretische Physik, 22761 Hamburg, Germany}

\begin{abstract}
Flavor-dependent neutrino transport is described by a well-known kinetic equation for occupation-number matrices in flavor space. However, in the context of fast flavor conversion, we identify an unforeseen predicament: the pivotal self-induced exponential growth of small inhomogeneities strongly violates conservation of neutrino-neutrino refractive energy. We prove that it is traded with the huge reservoir of neutrino kinetic energy through gradients of neutrino flavor coherence (the off-diagonal piece of the flavor density matrix) and derive the missing gradient terms. The usual equations remain sufficient to describe flavor evolution, at the cost of renouncing energy conservation, which cannot play any role in explaining the numerically observed final state.
\end{abstract}

\date{8 January 2024, revised 9 April 2024}

\maketitle

{\bf\textit{Introduction.}}---Flavor evolution in a dense neutrino environment is often described by a kinetic equation for the flavor density matrices $\rho(\bp,\bx,t)$, where the diagonal entries are the occupation numbers \cite{Dolgov:1980cq, Rudsky, Sigl:1993ctk, Sirera:1998ia, Yamada:2000za, Vlasenko:2013fja, Cirigliano:2014aoa, Volpe:2013uxl, Serreau:2014cfa, Kartavtsev:2015eva, Fiorillo:2024wej}. In the homogeneous case, they are defined as the expectation values of the number-operator matrices \smash{${\cal D}_{\alpha\beta}=a^\dagger_{\beta,\bp}a_{\alpha,\bp}$} with the flavor indices $\alpha$ and $\beta$ and $a^\dagger_{\alpha,\bp}$ the creation operator for a left-handed neutrino of flavor $\alpha$ with momentum $\bp$. In the limit of vanishing neutrino masses and flavor mixing, nontrivial flavor evolution can still arise through classical instabilities that would engender strong flavor correlations \cite{Sawyer:2015dsa, Chakraborty:2016lct, Izaguirre:2016gsx, Tamborra:2020cul, Richers:2022zug}. Dubbed {\em fast flavor evolution}, this subject has been intensely studied in view of practical consequences for neutrino flavor transport in stellar core collapse and binary neutron star mergers \cite{Xiong:2022vsy, Fernandez:2022yyv, Ehring:2023abs, Ehring:2023lcd, Nagakura:2023mhr, Nagakura:2023xhc, Zaizen:2023wht, Grohs:2023pgq, Abbar:2023zkm, Akaho:2023brj, Froustey:2023skf, Nagakura:2023wbf}.

Predicting the consequences of fast conversions is thus a central question. In tackling this problem, a key role must be played by conserved quantities, which are the only exact guide into the final state reached after conversions. One such quantity must be energy. However,we show that the usual equations of motion (EOMs) do not conserve energy.
In their often-used form, they are
\begin{equation}\label{eq:EOM}
    \bigl(\partial_t+\bv\cdot{\bm\nabla}_\br\bigr)\rho_{\bp,\br,t}
    =-i\bigl[{\sf H}_{\bp,\br,t},\rho_{\bp,\br,t}\bigr],
\end{equation}
where \smash{${\sf H}_{\bp,\br,t}=\sqrt{2}\GF\sum_{\bp'}\rho_{\bp',\br,t}\,(1-\cos\theta_{\bp,\bp'})$} is a matrix driving flavor evolution by neutrino-neutrino refraction.  The advection term proportional to $\bv=\bp/|\bp|$ quantifies a drift in coordinate space caused by inhomogeneities. This term allows for $\bp$-dependent instabilities, so even an initially homogeneous system can develop flavor disturbances growing exponentially and drifting to ever smaller scales~\cite{Duan:2014gfa, Mirizzi:2015fva, Bhattacharyya:2020jpj, Johns:2020qsk, Tamborra:2020cul, Richers:2021nbx, Bhattacharyya:2022eed}.

As detailed later, even an elementary example of two colliding beams consisting initially of $\nu_e$ and $\nu_x$ reveals that the refractive interaction energy changes dramatically as soon as initial seeds of inhomogeneity grow. The solution of this puzzle is that one cannot look at flavor evolution in isolation. Inhomogeneities in the weak potential exert a force and trade refractive interaction energy with the huge reservoir of neutrino kinetic energy. We here derive the missing terms in Eq.~\eqref{eq:EOM} and show that to lowest order in a gradient expansion, the energy exchange can be exactly accounted for. 

{\bf\textit{Interaction energy.}}---We first define the $\nu$--$\nu$ interaction energy. The starting point is the many-body Hamiltonian $\mathcal{H}=\mathcal{H}_0+\mathcal{U}$, where \smash{$\mathcal{H}_0=\sum_{\alpha,\bp}\epsilon_\bp a^\dagger_{\alpha,\bp}a_{\alpha,\bp}$} is the kinetic energy operator with $\epsilon_\bp=|\bp|$ in the massless limit. Moreover, the interaction energy is 
\begin{eqnarray}
    \mathcal{U}&=&\frac{\sqrt{2} \GF}{8}\sum_{\left\{\bp\right\},\alpha,\beta} a^\dagger_{\alpha,\bp_1}a_{\alpha,\bp_2}a^\dagger_{\beta,\bp_3}a_{\beta,\bp_4}\nonumber\\[-1ex]
    &&\kern7em{}\times\overline{u}_{\bp_1}\gamma^\mu u_{\bp_2}\overline{u}_{\bp_3}\gamma_\mu u_{\bp_4},
\end{eqnarray}
where $\GF$ is Fermi's constant, \smash{$\sum_{\left\{\bp\right\}}$} is performed over all momenta such that $\bp_1+\bp_3=\bp_2+\bp_4$, and $u_\bp$ is the spinor of a left-handed massless neutrino with momentum $\bp$, normalized such that $\overline{u}_\bp \gamma^0 u_\bp=2$. 

The energy of the system is the expectation value of $\mathcal{H}$ over the quantum state. In particular, the kinetic energy is $K=\langle \mathcal{H}_0 \rangle=\int d^3\br \sum_\bp \epsilon_\bp \mathrm{Tr}(\rho_\bp)$. For the interaction energy, we need to determine the average value of a string of four creation and annihilation operators. We use mean field approximation, where the quantum state is assumed to be the product of single-particle states. The expectation value factorizes as $\langle a^\dagger_1 a_2 a^\dagger_3 a_4\rangle=\langle a^\dagger_1 a_2\rangle \langle a^\dagger_3 a_4\rangle+(\delta_{2,3}-\langle a^\dagger_3 a_2\rangle) \langle a^\dagger_1 a_4\rangle$. In this way we define a potential energy of interaction $U=\langle \mathcal{U}\rangle$. For a homogeneous system, it is
\begin{eqnarray}\label{eq:Uint}
    U&=&\frac{\sqrt{2}\GF}{2}\int d^3\br\sum_{\bp,\bp'}
     \Bigl[\mathrm{Tr}(\rho_\bp)\,\mathrm{Tr}(\rho_{\bp'})+\mathrm{Tr}(\rho_\bp \rho_{\bp'})\Bigr]
     \nonumber\\[-1ex]
     &&\kern8em{}\times(1-\cos\theta_{\bp,\bp'}),
\end{eqnarray}
where we neglect an irrelevant (infinite) renormalization of the neutrino chemical potential. 

In the two-flavor case, the density matrices are often decomposed into their trace and trace-free part as
\begin{equation}
    \rho_\bp=\frac{P_\bp^0\,\mathbb{1}+\vP_\bp\cdot{\vec\sigma}}{2},
\end{equation}
where $\mathbb{1}$ is a unit matrix, $P_\bp^0=\mathrm{Tr}(\rho_{\bp})$, $\vec\sigma$ a vector of Pauli matrices, and \smash{$\vec P_\bp$} a polarization vector.
(\smash{$\vec P$} is a vector in flavor space, $\bp$ one in phase space.) Exactly one particle in mode $\bp$ implies \smash{$P_\bp^0=|\vec P_\bp|=1$}. Only \smash{$\vec P_\bp$} evolves by oscillations, whereas the trace (total particle number in a $\bp$-mode) is conserved. The oscillatory part of the interaction energy~is
\begin{equation}\label{eq:oscillation_energy}
    U_\mathrm{osc}=\frac{\sqrt{2}\GF}{4}\int d^3\br \sum_{\bp,\bp'}
    (1-\cos\theta_{\bp,\bp'})\,\vP_\bp\cdot\vP_{\bp'},
\end{equation}
corresponding to that part of Eq.~\eqref{eq:Uint} that depends on the trace-free parts of the $\rho$ matrices.

For simplicity we assume in the following a system that remains homogeneous in all but the spatial $z$-direction and use $v=v_z$. The standard EOM Eq.~\eqref{eq:EOM} falls into pieces for particle number and flavor polarization as
\begin{subequations}\label{eq:P-equations}
\begin{eqnarray}
    \partial_t P^0_\bp+\,v\partial_z P^0_\bp&=&0,
    \label{eq:trace}\\
    \partial_t \vec P_\bp+v\partial_z\vec P_\bp&=&
    \sqrt{2}\GF\bigl(\vP_0-v\vP_1\bigr)\times \vP_\bp,
    \label{eq:polarization}
\end{eqnarray}
\end{subequations}
where we use the angular moments \smash{$\vP_0=\sum_\bp \vP_\bp$} and \smash{$\vP_1=\sum_\bp v \vP_\bp$}. Thus, the usual equations decouple the neutrino number from their flavor evolution. 

Different from a neutrino gas subject to collisions with matter (see, e.g., Refs.~\cite{Johns:2021qby,Nagakura:2022qko, Xiong:2022zqz, Liu:2023pjw, Lin:2022dek, Johns:2022yqy, Shalgar:2022rjj,Padilla-Gay:2022wck,Fiorillo:2023ajs,Johns:2023xae}), where energy is not conserved anyway,  Eqs.~\eqref{eq:P-equations} describe a closed system that admits a conserved energy. The kinetic part $K=\int d^3\br \sum_\bp \epsilon_\bp P^0_\bp$ is seen to be conserved from Eq.~\eqref{eq:trace}. Also the nonoscillatory part of the interaction energy is separately conserved. $U_\mathrm{osc}=(\sqrt{2}\GF/4)\int d^3\br\,(\vP_0^2-\vP_1^2)$, which follows from Eq.~\eqref{eq:oscillation_energy}, must then be conserved as well. In the homogeneous case, this is indeed the case; the dynamics is periodic and described by a fictitious pendulum whose conserved energy coincides with $U_\mathrm{osc}$ \cite{Johns:2019izj, Padilla-Gay:2021haz, Fiorillo:2023mze, Fiorillo:2023hlk}. However, Eq.~\eqref{eq:polarization} immediately reveals that the conservation breaks down for inhomogeneous settings
\begin{equation}\label{eq:osc_energy_change}
    \frac{dU_\mathrm{osc}}{dt}=-\frac{\sqrt{2}\GF}{2}\int d^3\br\,
    \Bigl(\vP_0\cdot\partial_z \vP_1-\vP_1\cdot\partial_z \vP_2\Bigr),
\end{equation}
where $\vP_2=\sum_\bp v^2 \vP_\bp$. Therefore, the traditional EOMs are not consistent. 

{\bf{\textit{Two-beam example.}}}---The nonconservation of $U_\mathrm{osc}$ is a large effect as we show in a simple example of two opposite beams along the $z$-axis ($v=\pm 1$). The energies and number densities are equal, one beam initially occupied with $\nu_e$, the other with $\nu_x$. As usual, the beams represent many neutrinos with $\bp$'s close enough that the small spread is unimportant on the relevant time scales. Therefore, we represent the flavor polarization of each beam by a highly occupied $\rho$ matrix and concomitant polarization vector \smash{$\vP_{\pm}(z,t)$}. We thus need to solve
\begin{subequations}
    \begin{eqnarray}
      (\partial_t+\partial_z)\vP_+&=&2\,\vP_-\times\vP_+,
      \\
      (\partial_t-\partial_z)\vP_-&=&2\,\vP_+\times\vP_-,
    \end{eqnarray}
\end{subequations}
where the interaction strength was absorbed in the units of time and space. For \smash{$\partial_z\vP_\pm=0$} there is no instability, but the system has unstable inhomogeneous modes. 

We can show energy nonconservation analytically in the initial linear regime. In this limit, it is convenient to express the $x$--$y$ part of the polarization vector (the off-diagonal element of the density matrix) in the form $\psi_\pm=P^x_\pm+iP^y_\pm$ and initially we take $P_\pm^z(z,0)=\zeta_\pm$ to be constant. The linearized EOMs ($|\psi|\ll |\zeta|$)~are
\begin{subequations}\label{eq:EOM-linear-2}
    \begin{eqnarray}
     \bigl(\partial_t+\partial_z\bigr)\,\psi_+&=&2\,i\bigl(+\zeta_-\psi_+-\zeta_+\psi_-\bigr),
      \\
      \bigl(\partial_t-\partial_z\bigl)\,\psi_-&=&2\,i\bigl(-\zeta_-\psi_++\zeta_+\psi_-\bigr).
    \end{eqnarray}
\end{subequations}
These EOMs are most easily solved for the spatial Fourier modes $\widetilde\psi_\pm(k,t)=L^{-1}\int_{-L/2}^{+L/2} dz\,\psi_\pm(z,t)\,e^{-i k z}$ for a periodic box of length $L$ where $k=2\pi n/L$ with integer $n$ is a discrete wave vector. Equation~\eqref{eq:EOM-linear-2} shows that a $k$ mode is unstable if $k_1<k<k_2$ with $k_{1,2}=\zeta_--\zeta_+\mp2\sqrt{-\zeta_-\zeta_+}$. The corresponding eigenfrequencies are $\omega_k=\omega_0\pm i\gamma_k$, where the precession frequency $\omega_0=-\zeta_--\zeta_+$ does not depend on $k$ and the growth rate is $\gamma_k=\sqrt{(k-k_1)(k_2-k)}$. In particular, the maximum growth rate is attained for $\overline{k}=\zeta_--\zeta_+=(k_1+k_2)/2$ and is $\overline{\gamma}=2\sqrt{-\zeta_+ \zeta_-}=(k_2-k_1)/2$.

The oscillation energy $U_\mathrm{osc}=\int dz\,\vP_+\cdot\vP_-$ has one piece from the initial $z$ component, $U_\mathrm{osc}(0)=\zeta_+\zeta_- L$ that is conserved in the linear regime, and a piece from the off-diagonal terms $\Delta U_\mathrm{osc}=\int dz\,(\psi_+\psi_-^*+\psi_+^*\psi_-)/2=(L/2)\,\sum_k (\widetilde\psi_{+,k}\widetilde\psi_{-,k}^*+\widetilde\psi_{+,k}^*\widetilde\psi_{-,k})$, where in the continuous limit $\sum_k=\int dk/2\pi$. This piece can grow exponentially if there are small seeds for the unstable Fourier modes. If we decompose the $v=\pm1$ modes in terms of the eigenmodes of the linear analysis, the $k$ modes evolve~as
\begin{eqnarray}
 \kern-2em   \begin{pmatrix}
        \widetilde{\psi}_{+,k}(t)\\ 
        \widetilde{\psi}_{-,k}(t)
    \end{pmatrix}
    &=&\alpha_k\begin{pmatrix}
        k+\omega_0+i\gamma_k\\
        k-\omega_0-i\gamma_k
    \end{pmatrix}e^{(\gamma_k-i\omega_0)t}
    \nonumber\\
&&{}+\beta_k\begin{pmatrix}
        k+\omega_0-i\gamma_k\\ 
        k-\omega_0+i\gamma_k
    \end{pmatrix}e^{-(\gamma_k+i\omega_0)t}.
\end{eqnarray}
Matching this expression with the initial conditions $\widetilde{\psi}_{\pm,k}(0)=\widetilde{\psi}_{\pm,k}^0$ reveals for the growing piece
\begin{equation}
    \alpha_k=\frac{(\gamma_k+i\omega_0)\left(\widetilde{\psi}_{+,k}^0+\widetilde{\psi}_{-,k}^0\right)
    -i k \left(\widetilde{\psi}_{+,k}^0-\widetilde{\psi}_{-,k}^0\right)}{4k\gamma_k}.
\end{equation}
Substituting in the transverse part of the oscillation energy, we finally find
\begin{equation}
    \Delta U_\mathrm{osc}(t)=L\sum_{k=k_1}^{k_2}2k(k-\overline{k})|\alpha_k|^2 e^{2\gamma_k t},
\end{equation}
where we restrict the summation to the range of unstable eigenmodes. Notice that the most unstable eigenmode $k=\overline{k}$ does not contribute to energy nonconservation due to the vanishing prefactor. Otherwise, an exponential growth appears already at the linear level.

For more than two beams, from Eq.~\eqref{eq:osc_energy_change} it follows that $dU_\mathrm{osc}/dt\propto \sum_k k v \vP_{v}(k)\cdot \vP_{v'}^*(k)(1-vv')$, and thus it is still true that the energy change grows with $e^{2\overline{\gamma} t}$, provided that $k\neq 0$; homogeneous modes, with $k=0$, do not lead to energy change due to the prefactor $kv$.

For a numerical solution that extends to the nonlinear regime, we use periodic boundary conditions on a box of length $L=100$, divided in $N$ spatial bins. As initial conditions, we use $P^z_\pm(0)=\zeta_\pm=\pm1/2$. These parameters imply that the range of unstable modes is delimited by $k_1=-2$ and $k_2=0$, the maximum growth rate obtains for $\overline k=-1$ and is $\overline\gamma=1$, and the initial oscillation energy is $U_\mathrm{osc}(0)=\zeta_-\zeta_+ L=-25$. The transverse components $P^{x,y}_\pm$ are seeded with randomly sampled functions with a spatial dependence
\begin{equation}
    P^{x,y}_\pm=\sum_{n=-N_\mathrm{max}}^{N_\mathrm{max}}c^{x,y}_{\pm,n}e^{i\phi^{x,y}_{\pm,n}+i\frac{2\pi n z}{L}},
\end{equation}
where the amplitudes are sampled from a normal distribution with a variance $\sigma^2=10^{-8}$ and the phases are uniformly sampled from $0$ to $2\pi$; the functions are real and so $c_{n}=c_{-n}$ and $\phi_{n}=-\phi_{-n}$. We use arbitrarily $N_\mathrm{max}=100$ to avoid seeds at too small scales. Numerical stability requires keeping fluctuations at the scale of the grid spacing as small as possible.

\begin{figure}[t]
    \centering
    \includegraphics[width=1.0\columnwidth]{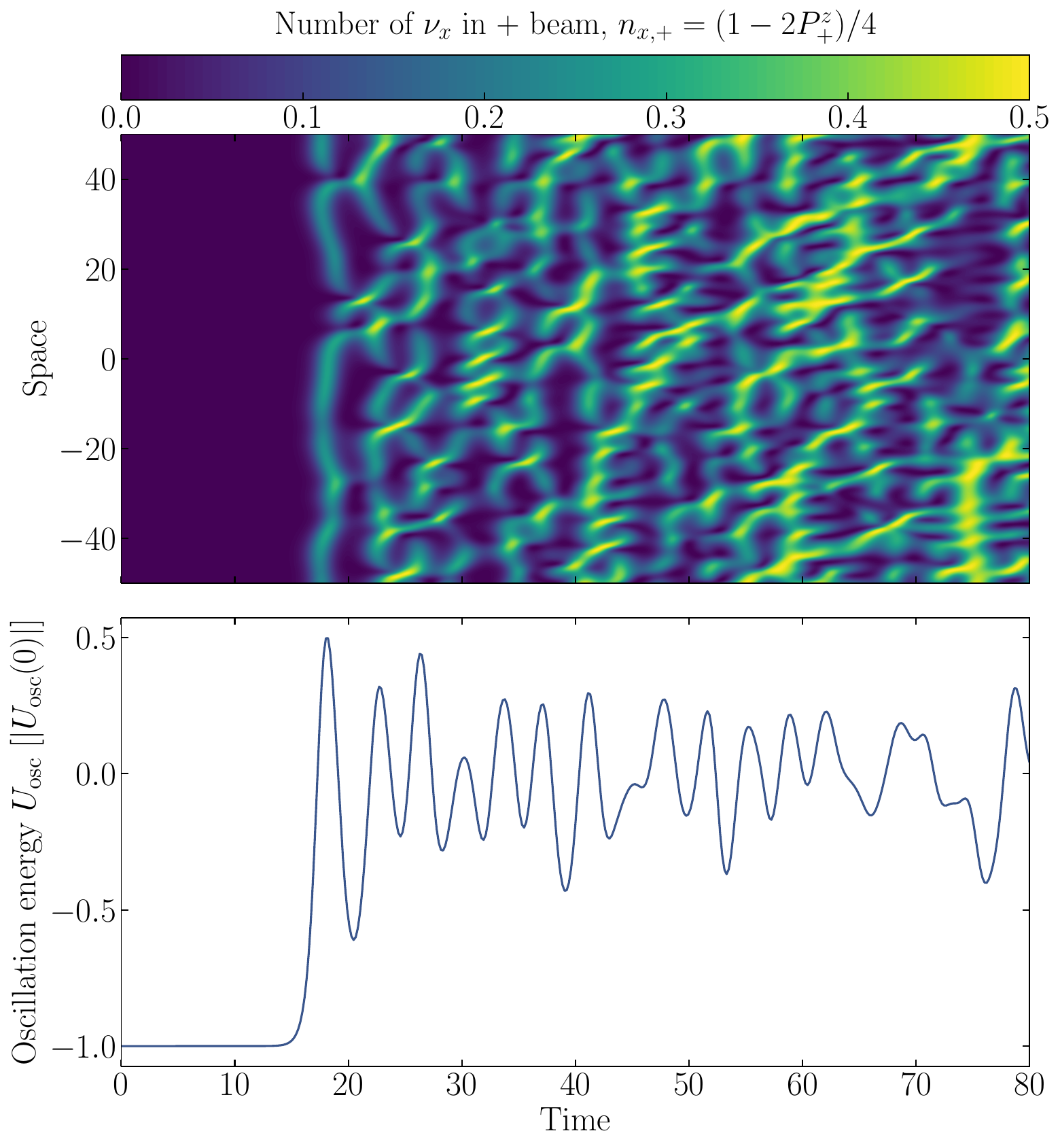}
    \caption{Solution of our two-beam example for one realization of initial seeds. {\em Top:} Contours of $\nu_x$ content of $v=+1$ beam as a function of $z$ and $t$. {\em Bottom:} Evolution of oscillation energy $U_\mathrm{osc}(t)$ in units of $U_\mathrm{osc}(0)$.}
    \label{fig:two-beams}
\end{figure}

Figure~\ref{fig:two-beams} shows the solution for a single realization of initial seeds. The upper panel shows contours of the $\nu_x$-content of the beam initially occupied with $\nu_e$, as a function of $z$ and $t$. The lower panel shows $U_\mathrm{osc}(t)$ in units of $U_\mathrm{osc}(0)$. On the shown timescale, the range of unstable modes $-2<k<0$ grows nonlinear and then keeps oscillating, with beats between different modes causing an irregular pattern that in detail depends on the choice of seeds. In the longer run, these nonlinear modes feed higher-$k$ modes and flavor variations will obtain on ever smaller scales. If we had used a larger box, or equivalently averaged over more than one realization of the initial conditions, the amplitude of the oscillations in the final state would shrink. Nevertheless, the bulk of the nonconservation of $U_\mathrm{osc}$ happens in the initial phase, and is clearly a large effect.

{\bf\textit{Inhomogeneous kinetic equations.}}---The lack of refractive-energy conservation questions the validity of the traditional EOMs. The strategy for their derivation is a perturbative expansion in three small parameters: (i)~The mass-to-energy ratio $r_m=m_\nu/\epsilon_\bp$, where $m_\nu$ is the neutrino mass, but in our fast flavor limit we do not worry about it. (ii)~The fractional refractive energy shift $r_\mu=\mu/\epsilon_\bp$, where \smash{$\mu=\sqrt{2}\GF n_\nu$} is a scale for the $\nu$--$\nu$ interaction energy. For $T\simeq5$~MeV in the decoupling region of a supernova, $n_\nu\simeq 10^{33}$~cm$^{-3}$, $\mu\simeq0.2$~meV, and $r_\mu\simeq 10^{-11}$ and thus indeed very small. (iii)~The scale of density variations $\ell$ in units of the neutrino de Broglie wavelength $r_\ell=(\ell |\bp|)^{-1}\ll1$. Density variations are on scales much larger than $|\bp|^{-1}$. The EOMs are typically derived to lowest order in all of these small ratios, notably neglecting terms of the order of $r_\ell r_\mu$, which causes the nonconservation of energy.

To augment Eq.~\eqref{eq:EOM} with the missing terms, we note that in an inhomogeneous setting, the space-dependent occupation-number matrix is defined by a Wigner distribution \cite{Wigner:1932eb}, the expectation value of
\begin{equation}
    \mathcal{D}_{\alpha\beta}(\br,\bp)=\sum_\bk a^\dagger_{\beta,\bp+\bkk/2}a_{\alpha,\bp-\bkk/2}e^{-i\bk\cdot \br}.
\end{equation}
Here, the Fourier wavevector $\bk$ measures the typical length scale over which the density matrix is changing, and is therefore $|\bk|\sim\ell^{-1}\ll |\bp|$. The time derivative is obtained using Heisenberg's equations. The commutator with the kinetic energy is easily established as
\begin{eqnarray}
    \left(\frac{\partial\mathcal{D}_{\alpha\beta}}{\partial t}\right)_0&=& i\sum_\bk a^\dagger_{\beta,\bp+\frac{\bk}{2}}a_{\alpha,\bp-\frac{\bk}{2}}e^{-i\bk\cdot\br}
    \left(\epsilon_{\bp+\frac{\bk}{2}}-\epsilon_{\bp-\frac{\bk}{2}}\right)
    \nonumber\\
    &\simeq&-\bv\cdot{\bm\nabla}_\br \mathcal{D}_{\alpha\beta},
\end{eqnarray}
recovering the usual advection term. 

For the interaction part, we here outline the general strategy and report details in the Supplemental Material (SM). The commutator $[\mathcal{D},\mathcal{U}]$ produces strings of four creation and annihilation operators, or four-point correlation functions. The average of these combinations are evaluated again using the mean-field approximation. In the homogeneous limit, when taking the expectation value of $\langle a^\dagger_1 a_2\rangle$, the momenta of state $1$ and $2$ must be equal. We go one step further and include the inhomogeneity to first order in $\bk$. The final result, fully derived in the SM, is
\eject
\begin{widetext}
    \begin{equation}\label{eq:full_eom_schematic}
         \partial_t \rho_{\bp,\br}+\bv\cdot{\bm\nabla}_\br \rho_{\bp,\br}
         +\frac{1}{2}\left\{{\bm\nabla}_\bp {\sf\Omega}^{(0)}_{\bp,\br},{\bm\nabla}_\br \rho_{\bp,\br}\right\}
         -\frac{1}{2}\left\{{\bm\nabla}_\br {\sf\Omega}^{(0)}_{\bp,\br},{\bm\nabla}_\bp \rho_{\bp,\br}\right\}
    =i\Bigl[\rho_{\bp,\br},{\sf\Omega}_{\bp,\br}\Bigr],
    \end{equation}

where we introduce
\begin{equation}
{\sf\Omega}_{\bp,\br}^{(0)}=\sqrt{2}\GF\sum_{\bp'}(1-\cos\theta_{\bp,\bp'})
\Bigl[\rho_{\bp',\br}+\mathrm{Tr}(\rho_{\bp',\br})\mathbb{1}\Bigr]
\end{equation}
that coincides with the standard ${\sf H}_{\bp,\br}$ stated after Eq.~\eqref{eq:EOM} 
except for the additional trace term that drops out in commutators,
but not in the anticommutators on the left-hand side. Moreover, 
\begin{equation}\label{eq:Omega}
{\sf\Omega}_{\bp,\br}= {\sf\Omega}_{\bp,\br}^{(0)}+
\sqrt{2}\GF\sum_{\bp'}
    \frac{p+p'}{p^2p^{'2}}\,\left(\bp\times\bp'\right)\cdot\Bigl[{\bm\nabla}_\br \rho_{\bp',\br}
    +\mathrm{Tr}({\bm\nabla}_\br\rho_{\bp',\br})\mathbb{1} \Bigr],
\end{equation}
\end{widetext}
where $p=|\bp|$ and $p'=|\bp'|$. 

 ${\sf\Omega}_{\bp,\br}$ is the renormalized quasiparticle energy, in general a matrix in flavor space, analogous to the renormalized quasiparticle energy in Landau's theory of  a Fermi liquid~\cite{Landau:1958joj}. This induces a renormalization in the group velocity \smash{${\bm\nabla}_\bp {\sf\Omega}^{(0)}_{\bp,\br}$} (only the zero order terms must be kept which already contain one gradient) and a weak force field \smash{$-{\bm\nabla}_\br {\sf\Omega}^{(0)}_{\bp,\br}$}.

All additional terms are small, of the order of $r_\mu r_l$, and thus will not produce \textit{quantitatively} large flavor-conversion effects. However, as a \textit{qualitatively} new effect, the renormalized group velocity causes a slow spatial drift of neutrinos; the order of magnitude of the velocity is \smash{$|{\bm\nabla}_\bp {\sf\Omega}^{(0)}_{\bp,\br}|\sim r_\mu$}. Since $r_\mu\sim 10^{-11}$, this is completely negligible compared to the standard neutrino velocity.

The most significant impact comes from the weak force  \smash{$-{\bm\nabla}_\br {\sf\Omega}^{(0)}_{\bp,\br}$}, which can change the neutrino kinetic energy. Part of this force originates from gradients in the neutrino number density. Our new insight is that an additional component originates from gradients in flavor composition. To see this clearly, we rewrite the EOMs~as
\begin{widetext}
\begin{equation}
    \partial_t \rho_{\bp,\br}+\bv\cdot{\bm\nabla}_\br \rho_{\bp,\br}
    =i\bigl[\rho_{\bp,\br},{\sf\Omega}_{\bp,\br}\bigr]
    +{\bm\nabla}_\br\cdot {\bPhi}_{\bp,\br}-{\bm\nabla}_\bp\cdot {\bF}_{\bp,\br},
\end{equation}
where
\begin{equation}
    \bPhi_{\bp,\br}=\frac{\sqrt{2}\GF}{2}\sum_{\bp'}(1-\cos\theta_{\bp,\bp'})
    \Bigl[\bigl\{\rho_{\bp',\br},{\bm\nabla}_\bp\rho_{\bp,\br}\bigr\}
+2\,\mathrm{Tr}(\rho_{\bp',\br}){\bm\nabla}_\bp\rho_{\bp,\br}\Bigr]
\end{equation}
is the number flux of neutrinos passing through a surface element of coordinate space at phase-space location $\{\bp,\br\}$, as seen from the structure analogous to a continuity equation, whereas $\bF_{\bp,\br}$ is the same in momentum space, given by the same expression with ${\bm\nabla}_\bp\to{\bm\nabla}_\br$. This term couples the trace of the density matrix with the polarization vector; taking the trace, we find
\begin{equation}\label{eq:tr_flux}
\mathrm{Tr}(\bF_{\bp,\br})=\frac{\sqrt{2}\GF}{2}\sum_{\bp'}(1-\cos\theta_{\bp,\bp'})
    \Bigl[3P^0_{\bp',\br}{\bm\nabla}_\br P^0_{\bp,\br}
   +\vP_{\bp',\br}\cdot{\bm\nabla}_\br\vP_{\bp,\br}\Bigr].
\end{equation}
\end{widetext}
Thus, spatial gradients of the polarization vectors can feed $\nu$ kinetic energy. This term implies a net rate of energy gain or loss of order $d\epsilon_\bp/dt\simeq \mu/\ell$. Over a timescale $\ell$, the energy that a $\nu$ can accumulate is of order $\mu\ll\epsilon_\bp$, so we recover that this is a small effect \textit{relative} to $\epsilon_\bp$, but large relative to the interaction energy. It provides the missing channel by which kinetic and refractive energy can be traded. Even a small amount of energy $\mu$ lost (gained) from the large $\nu$ kinetic energy explains the large change in the refractive energy, since $U_\mathrm{osc}/K\simeq \mu/\epsilon_\bp$.

With the new kinetic equations, we can directly prove that to first order in $r_\mu$ and $r_\ell$, the total energy is conserved. 
After integrating by parts, we find
\begin{equation}\label{eq:kinetic_energy_evolution}
    \frac{dK}{dt}=\int d^3\br \sum_\bp \mathrm{Tr}\left(\bv\cdot\bF_{\bp,\br}\right).
\end{equation}
Substituting the polarization vector part of Eq.~\eqref{eq:tr_flux} in Eq.~\eqref{eq:kinetic_energy_evolution}, we find that it precisely balances the rate of change in the oscillation energy found in Eq.~\eqref{eq:osc_energy_change}.

{\bf\textit{Discussion.}}---We have shown that the usual kinetic equation for mixed neutrinos Eq.~\eqref{eq:EOM} is non-conservative in the inhomogeneous case. We have illustrated this point with a simple two-beam example, where the $\nu$--$\nu$ interaction energy strongly changes, and we have also shown this effect analytically in the linear regime. Deriving missing gradient terms beyond Eq.~\eqref{eq:EOM}, we have shown that the refractive energy gained or lost is precisely traded with neutrino kinetic energy that usually is not followed. In our two-beam example, the monochromatic initial energy distribution develops small space-time dependent shifts that account for the missing energy. We have not tried to study this effect numerically because it requires many $\bp$-modes around the original one, but we have proven energy conservation analytically.

The usual EOMs correctly account for flavor evolution, meaning the trace-free part of the density matrices often described by polarization vectors. On the other hand, the particle number in a given $\bp$-mode is conserved, corresponding to the trace part of the EOM, where the left-hand side of Eq.~\eqref{eq:EOM} is a continuity equation. Therefore, the trace part of the EOM must be expanded to higher order in the gradients to capture nontrivial evolution. The reshuffling of neutrinos among $\bp$-modes is a small effect relative to the large kinetic energies, yet precisely absorbs the missing refractive energy. 

These novel terms come from the gradients of the neutrino self-energy. With hindsight, that they would affect neutrino evolution, is obvious from the viewpoint of physical kinetics, and implicitly present in previous formal derivations \cite{Sigl:1993ctk,Rudsky, Yamada:2000za, Vlasenko:2013fja, Cirigliano:2014aoa}, but usually assumed to be a small effect. Our new insight is that fast conversions spontaneously break homogeneity, magnifying the gradients of the trace-free density matrix, making these terms large enough to explain completely the apparent non-conservation of energy.

We note that Eq.~\eqref{eq:EOM} conserves entropy and including the new gradient terms, this is also the case, i.e., entropy is conserved order by order in the gradient expansion (see Supplemental Material). On the practical level, the conservation of entropy, but not of energy, could be used to test numerical stability. Even more importantly, in making predictions on the final state induced by conversions, energy conservation cannot be used in practice, unless the new terms are kept.
 
Our finding may shed new light on a recent proposal that the final outcome of fast conversions may be some sort of thermalized state~\cite{Johns:2023jjt}. The quasi-steady state generically observed in numerical simulations of fast conversions~\cite{Padilla-Gay:2020uxa,Zaizen:2023ihz, Xiong:2023vcm, Abbar:2023ltx, Cornelius:2023eop, Urquilla:2024bvf} is only determined by the oscillatory part of the density matrix, yet its dynamics does not admit a separately conserved energy. Thus, it cannot separately thermalize, since it exchanges energy with the kinetic energy of neutrinos.

{\bf\textit{Acknowledgments.}}---We acknowledge useful discussion with Basudeb Dasgupta, Lucas Johns, Shashank Shalgar, and Irene Tamborra. DFGF is supported by the Villum Fonden under Project No.\ 29388 and the European Union's Horizon 2020 Research and Innovation Program under the Marie Sk{\l}odowska-Curie Grant Agreement No.\ 847523 ``INTERACTIONS.'' GGR acknowledges partial support by the German Research Foundation (DFG) through the Collaborative Research Centre ``Neutrinos and Dark Matter in Astro- and Particle Physics (NDM),'' Grant SFB-1258-283604770, and under Germany’s Excellence Strategy through the Cluster of Excellence ORIGINS EXC-2094-390783311. GS acknowledges support by the Deutsche Forschungsgemeinschaft (DFG, German Research Foundation) under Germany’s Excellence Strategy -- EXC 2121 ``Quantum Universe'' -- 390833306.

\bibliographystyle{bibi}
\bibliography{Biblio}

\providecommand{\href}[2]{#2}\begingroup\raggedright\begin{thebibliography}{10}

\bibitem{Dolgov:1980cq}
A.~D. Dolgov, \emph{{Neutrinos in the early universe}}, {\emph{Sov. J. Nucl.
  Phys.} {\bfseries 33} (1981) 700}. [{\em Yad.\ Fiz.} {\bf 33} (1981) 1309].

\bibitem{Rudsky}
M.~A. {Rudzsky}, \emph{{Kinetic equations for neutrino spin- and
  type-oscillations in a medium}},
  \href{https://doi.org/10.1007/BF00653658}{\emph{Astrophys. Space Sci}
  {\bfseries 165} (1990) 65}.

\bibitem{Sigl:1993ctk}
G.~Sigl and G.~Raffelt, \emph{{General kinetic description of relativistic
  mixed neutrinos}},
  \href{https://doi.org/10.1016/0550-3213(93)90175-O}{\emph{Nucl. Phys. B}
  {\bfseries 406} (1993) 423}.

\bibitem{Sirera:1998ia}
M.~Sirera and A.~P{\'e}rez, \emph{{Relativistic Wigner function approach to
  neutrino propagation in matter}},
  \href{https://doi.org/10.1103/PhysRevD.59.125011}{\emph{Phys. Rev. D}
  {\bfseries 59} (1999) 125011}
  [\href{https://arxiv.org/abs/hep-ph/9810347}{{\ttfamily hep-ph/9810347}}].

\bibitem{Yamada:2000za}
S.~Yamada, \emph{{Boltzmann equations for neutrinos with flavor mixings}},
  \href{https://doi.org/10.1103/PhysRevD.62.093026}{\emph{Phys. Rev. D}
  {\bfseries 62} (2000) 093026}
  [\href{https://arxiv.org/abs/astro-ph/0002502}{{\ttfamily
  astro-ph/0002502}}].

\bibitem{Vlasenko:2013fja}
A.~Vlasenko, G.~M. Fuller and V.~Cirigliano, \emph{{Neutrino Quantum
  Kinetics}}, \href{https://doi.org/10.1103/PhysRevD.89.105004}{\emph{Phys.
  Rev. D} {\bfseries 89} (2014) 105004}
  [\href{https://arxiv.org/abs/1309.2628}{{\ttfamily 1309.2628}}].

\bibitem{Cirigliano:2014aoa}
V.~Cirigliano, G.~M. Fuller and A.~Vlasenko, \emph{{A New Spin on Neutrino
  Quantum Kinetics}},
  \href{https://doi.org/10.1016/j.physletb.2015.04.066}{\emph{Phys. Lett. B}
  {\bfseries 747} (2015) 27} [\href{https://arxiv.org/abs/1406.5558}{{\ttfamily
  1406.5558}}].

\bibitem{Volpe:2013uxl}
C.~Volpe, D.~V\"a\"an\"anen and C.~Espinoza, \emph{{Extended evolution
  equations for neutrino propagation in astrophysical and cosmological
  environments}}, \href{https://doi.org/10.1103/PhysRevD.87.113010}{\emph{Phys.
  Rev. D} {\bfseries 87} (2013) 113010}
  [\href{https://arxiv.org/abs/1302.2374}{{\ttfamily 1302.2374}}].

\bibitem{Serreau:2014cfa}
J.~Serreau and C.~Volpe, \emph{{Neutrino-antineutrino correlations in dense
  anisotropic media}},
  \href{https://doi.org/10.1103/PhysRevD.90.125040}{\emph{Phys. Rev. D}
  {\bfseries 90} (2014) 125040}
  [\href{https://arxiv.org/abs/1409.3591}{{\ttfamily 1409.3591}}].

\bibitem{Kartavtsev:2015eva}
A.~Kartavtsev, G.~Raffelt and H.~Vogel, \emph{{Neutrino propagation in media:
  Flavor, helicity, and pair correlations}},
  \href{https://doi.org/10.1103/PhysRevD.91.125020}{\emph{Phys. Rev. D}
  {\bfseries 91} (2015) 125020}
  [\href{https://arxiv.org/abs/1504.03230}{{\ttfamily 1504.03230}}].

\bibitem{Fiorillo:2024wej}
D.~F.~G. Fiorillo, G.~G. Raffelt and G.~Sigl, \emph{{Collective
  neutrino-antineutrino oscillations in dense neutrino environments?}},
  \href{https://doi.org/10.1103/PhysRevD.109.043031}{\emph{Phys. Rev. D}
  {\bfseries 109} (2024) 043031}
  [\href{https://arxiv.org/abs/2401.02478}{{\ttfamily 2401.02478}}].

\bibitem{Sawyer:2015dsa}
R.~F. Sawyer, \emph{{Neutrino cloud instabilities just above the neutrino
  sphere of a supernova}},
  \href{https://doi.org/10.1103/PhysRevLett.116.081101}{\emph{Phys. Rev. Lett.}
  {\bfseries 116} (2016) 081101}
  [\href{https://arxiv.org/abs/1509.03323}{{\ttfamily 1509.03323}}].

\bibitem{Chakraborty:2016lct}
S.~Chakraborty, R.~S. Hansen, I.~Izaguirre and G.~Raffelt, \emph{{Self-induced
  neutrino flavor conversion without flavor mixing}},
  \href{https://doi.org/10.1088/1475-7516/2016/03/042}{\emph{JCAP} {\bfseries
  03} (2016) 042} [\href{https://arxiv.org/abs/1602.00698}{{\ttfamily
  1602.00698}}].

\bibitem{Izaguirre:2016gsx}
I.~Izaguirre, G.~Raffelt and I.~Tamborra, \emph{{Fast Pairwise Conversion of
  Supernova Neutrinos: A Dispersion-Relation Approach}},
  \href{https://doi.org/10.1103/PhysRevLett.118.021101}{\emph{Phys. Rev. Lett.}
  {\bfseries 118} (2017) 021101}
  [\href{https://arxiv.org/abs/1610.01612}{{\ttfamily 1610.01612}}].

\bibitem{Tamborra:2020cul}
I.~Tamborra and S.~Shalgar, \emph{{New Developments in Flavor Evolution of a
  Dense Neutrino Gas}},
  \href{https://doi.org/10.1146/annurev-nucl-102920-050505}{\emph{Ann. Rev.
  Nucl. Part. Sci.} {\bfseries 71} (2021) 165}
  [\href{https://arxiv.org/abs/2011.01948}{{\ttfamily 2011.01948}}].

\bibitem{Richers:2022zug}
S.~Richers and M.~Sen, \emph{{Fast Flavor Transformations}},
  \href{https://doi.org/10.1007/978-981-15-8818-1_125-1}{\emph{Handbook of
  Nuclear Physics} (2022) 1}
  [\href{https://arxiv.org/abs/2207.03561}{{\ttfamily 2207.03561}}].

\bibitem{Xiong:2022vsy}
Z.~Xiong, M.-R. Wu, G.~Mart\'\i{}nez-Pinedo, T.~Fischer, M.~George, C.-Y. Lin
  and L.~Johns, \emph{{Evolution of collisional neutrino flavor instabilities
  in spherically symmetric supernova models}},
  \href{https://doi.org/10.1103/PhysRevD.107.083016}{\emph{Phys. Rev. D}
  {\bfseries 107} (2023) 083016}
  [\href{https://arxiv.org/abs/2210.08254}{{\ttfamily 2210.08254}}].

\bibitem{Fernandez:2022yyv}
R.~Fern\'andez, S.~Richers, N.~Mulyk and S.~Fahlman, \emph{{Fast flavor
  instability in hypermassive neutron star disk outflows}},
  \href{https://doi.org/10.1103/PhysRevD.106.103003}{\emph{Phys. Rev. D}
  {\bfseries 106} (2022) 103003}
  [\href{https://arxiv.org/abs/2207.10680}{{\ttfamily 2207.10680}}].

\bibitem{Ehring:2023abs}
J.~Ehring, S.~Abbar, H.-T. Janka, G.~Raffelt and I.~Tamborra, \emph{{Fast
  Neutrino Flavor Conversions Can Help and Hinder Neutrino-Driven Explosions}},
  \href{https://doi.org/10.1103/PhysRevLett.131.061401}{\emph{Phys. Rev. Lett.}
  {\bfseries 131} (2023) 061401}
  [\href{https://arxiv.org/abs/2305.11207}{{\ttfamily 2305.11207}}].

\bibitem{Ehring:2023lcd}
J.~Ehring, S.~Abbar, H.-T. Janka, G.~Raffelt and I.~Tamborra, \emph{{Fast
  neutrino flavor conversion in core-collapse supernovae: A parametric study in
  1D models}}, \href{https://doi.org/10.1103/PhysRevD.107.103034}{\emph{Phys.
  Rev. D} {\bfseries 107} (2023) 103034}
  [\href{https://arxiv.org/abs/2301.11938}{{\ttfamily 2301.11938}}].

\bibitem{Nagakura:2023mhr}
H.~Nagakura, \emph{{Roles of Fast Neutrino-Flavor Conversion on the
  Neutrino-Heating Mechanism of Core-Collapse Supernova}},
  \href{https://doi.org/10.1103/PhysRevLett.130.211401}{\emph{Phys. Rev. Lett.}
  {\bfseries 130} (2023) 211401}
  [\href{https://arxiv.org/abs/2301.10785}{{\ttfamily 2301.10785}}].

\bibitem{Nagakura:2023xhc}
H.~Nagakura and M.~Zaizen, \emph{{Basic characteristics of neutrino flavor
  conversions in the postshock regions of core-collapse supernova}},
  \href{https://doi.org/10.1103/PhysRevD.108.123003}{\emph{Phys. Rev. D}
  {\bfseries 108} (2023) 123003}
  [\href{https://arxiv.org/abs/2308.14800}{{\ttfamily 2308.14800}}].

\bibitem{Zaizen:2023wht}
M.~Zaizen and H.~Nagakura, \emph{{Fast neutrino-flavor swap in high-energy
  astrophysical environments}},
  \href{https://doi.org/10.1103/PhysRevD.109.083031}{\emph{Phys. Rev. D}
  {\bfseries 109} (2024) 083031}
  [\href{https://arxiv.org/abs/2311.13842}{{\ttfamily 2311.13842}}].

\bibitem{Grohs:2023pgq}
E.~Grohs, S.~Richers, S.~M. Couch, F.~Foucart, J.~Froustey, J.~P. Kneller and
  G.~C. McLaughlin, \emph{{Two-moment Neutrino Flavor Transformation with
  Applications to the Fast Flavor Instability in Neutron Star Mergers}},
  \href{https://doi.org/10.3847/1538-4357/ad13f2}{\emph{Astrophys. J.}
  {\bfseries 963} (2024) 11}
  [\href{https://arxiv.org/abs/2309.00972}{{\ttfamily 2309.00972}}].

\bibitem{Abbar:2023zkm}
S.~Abbar and H.~Nagakura, \emph{{Detecting fast neutrino flavor conversions
  with machine learning}},
  \href{https://doi.org/10.1103/PhysRevD.109.023033}{\emph{Phys. Rev. D}
  {\bfseries 109} (2024) 023033}
  [\href{https://arxiv.org/abs/2310.03807}{{\ttfamily 2310.03807}}].

\bibitem{Akaho:2023brj}
R.~Akaho, J.~Liu, H.~Nagakura, M.~Zaizen and S.~Yamada, \emph{{Collisional and
  fast neutrino flavor instabilities in two-dimensional core-collapse supernova
  simulation with Boltzmann neutrino transport}},
  \href{https://doi.org/10.1103/PhysRevD.109.023012}{\emph{Phys. Rev. D}
  {\bfseries 109} (2024) 023012}
  [\href{https://arxiv.org/abs/2311.11272}{{\ttfamily 2311.11272}}].

\bibitem{Froustey:2023skf}
J.~Froustey, S.~Richers, E.~Grohs, S.~D. Flynn, F.~Foucart, J.~P. Kneller and
  G.~C. McLaughlin, \emph{{Neutrino fast flavor oscillations with moments:
  Linear stability analysis and application to neutron star mergers}},
  \href{https://doi.org/10.1103/PhysRevD.109.043046}{\emph{Phys. Rev. D}
  {\bfseries 109} (2024) 043046}
  [\href{https://arxiv.org/abs/2311.11968}{{\ttfamily 2311.11968}}].

\bibitem{Nagakura:2023wbf}
H.~Nagakura, \emph{{Global features of fast neutrino-flavor conversion in
  binary neutron star mergers}},
  \href{https://doi.org/10.1103/PhysRevD.108.103014}{\emph{Phys. Rev. D}
  {\bfseries 108} (2023) 103014}
  [\href{https://arxiv.org/abs/2306.10108}{{\ttfamily 2306.10108}}].

\bibitem{Duan:2014gfa}
H.~Duan and S.~Shalgar, \emph{{Flavor instabilities in the neutrino line
  model}}, \href{https://doi.org/10.1016/j.physletb.2015.05.057}{\emph{Phys.
  Lett. B} {\bfseries 747} (2015) 139}
  [\href{https://arxiv.org/abs/1412.7097}{{\ttfamily 1412.7097}}].

\bibitem{Mirizzi:2015fva}
A.~Mirizzi, G.~Mangano and N.~Saviano, \emph{{Self-induced flavor instabilities
  of a dense neutrino stream in a two-dimensional model}},
  \href{https://doi.org/10.1103/PhysRevD.92.021702}{\emph{Phys. Rev. D}
  {\bfseries 92} (2015) 021702}
  [\href{https://arxiv.org/abs/1503.03485}{{\ttfamily 1503.03485}}].

\bibitem{Bhattacharyya:2020jpj}
S.~Bhattacharyya and B.~Dasgupta, \emph{{Fast Flavor Depolarization of
  Supernova Neutrinos}},
  \href{https://doi.org/10.1103/PhysRevLett.126.061302}{\emph{Phys. Rev. Lett.}
  {\bfseries 126} (2021) 061302}
  [\href{https://arxiv.org/abs/2009.03337}{{\ttfamily 2009.03337}}].

\bibitem{Johns:2020qsk}
L.~Johns, H.~Nagakura, G.~M. Fuller and A.~Burrows, \emph{{Fast oscillations,
  collisionless relaxation, and spurious evolution of supernova neutrino
  flavor}}, \href{https://doi.org/10.1103/PhysRevD.102.103017}{\emph{Phys. Rev.
  D} {\bfseries 102} (2020) 103017}
  [\href{https://arxiv.org/abs/2009.09024}{{\ttfamily 2009.09024}}].

\bibitem{Richers:2021nbx}
S.~Richers, D.~E. Willcox, N.~M. Ford and A.~Myers, \emph{{Particle-in-cell
  simulation of the neutrino fast flavor instability}},
  \href{https://doi.org/10.1103/PhysRevD.103.083013}{\emph{Phys. Rev. D}
  {\bfseries 103} (2021) 083013}
  [\href{https://arxiv.org/abs/2101.02745}{{\ttfamily 2101.02745}}].

\bibitem{Bhattacharyya:2022eed}
S.~Bhattacharyya and B.~Dasgupta, \emph{{Elaborating the ultimate fate of fast
  collective neutrino flavor oscillations}},
  \href{https://doi.org/10.1103/PhysRevD.106.103039}{\emph{Phys. Rev. D}
  {\bfseries 106} (2022) 103039}
  [\href{https://arxiv.org/abs/2205.05129}{{\ttfamily 2205.05129}}].

\bibitem{Johns:2021qby}
L.~Johns, \emph{{Collisional Flavor Instabilities of Supernova Neutrinos}},
  \href{https://doi.org/10.1103/PhysRevLett.130.191001}{\emph{Phys. Rev. Lett.}
  {\bfseries 130} (2023) 191001}
  [\href{https://arxiv.org/abs/2104.11369}{{\ttfamily 2104.11369}}].

\bibitem{Nagakura:2022qko}
H.~Nagakura, \emph{{General-relativistic quantum-kinetics neutrino transport}},
  \href{https://doi.org/10.1103/PhysRevD.106.063011}{\emph{Phys. Rev. D}
  {\bfseries 106} (2022) 063011}
  [\href{https://arxiv.org/abs/2206.04098}{{\ttfamily 2206.04098}}].

\bibitem{Xiong:2022zqz}
Z.~Xiong, L.~Johns, M.-R. Wu and H.~Duan, \emph{{Collisional flavor instability
  in dense neutrino gases}},
  \href{https://doi.org/10.1103/PhysRevD.108.083002}{\emph{Phys. Rev. D}
  {\bfseries 108} (2023) 083002}
  [\href{https://arxiv.org/abs/2212.03750}{{\ttfamily 2212.03750}}].

\bibitem{Liu:2023pjw}
J.~Liu, M.~Zaizen and S.~Yamada, \emph{{Systematic study of the resonancelike
  structure in the collisional flavor instability of neutrinos}},
  \href{https://doi.org/10.1103/PhysRevD.107.123011}{\emph{Phys. Rev. D}
  {\bfseries 107} (2023) 123011}
  [\href{https://arxiv.org/abs/2302.06263}{{\ttfamily 2302.06263}}].

\bibitem{Lin:2022dek}
Y.-C. Lin and H.~Duan, \emph{{Collision-induced flavor instability in dense
  neutrino gases with energy-dependent scattering}},
  \href{https://doi.org/10.1103/PhysRevD.107.083034}{\emph{Phys. Rev. D}
  {\bfseries 107} (2023) 083034}
  [\href{https://arxiv.org/abs/2210.09218}{{\ttfamily 2210.09218}}].

\bibitem{Johns:2022yqy}
L.~Johns and Z.~Xiong, \emph{{Collisional instabilities of neutrinos and their
  interplay with fast flavor conversion in compact objects}},
  \href{https://doi.org/10.1103/PhysRevD.106.103029}{\emph{Phys. Rev. D}
  {\bfseries 106} (2022) 103029}
  [\href{https://arxiv.org/abs/2208.11059}{{\ttfamily 2208.11059}}].

\bibitem{Shalgar:2022rjj}
S.~Shalgar and I.~Tamborra, \emph{{Neutrino decoupling is altered by flavor
  conversion}}, \href{https://doi.org/10.1103/PhysRevD.108.043006}{\emph{Phys.
  Rev. D} {\bfseries 108} (2023) 043006}
  [\href{https://arxiv.org/abs/2206.00676}{{\ttfamily 2206.00676}}].

\bibitem{Padilla-Gay:2022wck}
I.~Padilla-Gay, I.~Tamborra and G.~G. Raffelt, \emph{{Neutrino fast flavor
  pendulum. II. Collisional damping}},
  \href{https://doi.org/10.1103/PhysRevD.106.103031}{\emph{Phys. Rev. D}
  {\bfseries 106} (2022) 103031}
  [\href{https://arxiv.org/abs/2209.11235}{{\ttfamily 2209.11235}}].

\bibitem{Fiorillo:2023ajs}
D.~F.~G. Fiorillo, I.~Padilla-Gay and G.~G. Raffelt, \emph{{Collisions and
  collective flavor conversion: Integrating out the fast dynamics}},
  \href{https://doi.org/10.1103/PhysRevD.109.063021}{\emph{Phys. Rev. D}
  {\bfseries 109} (2024) 063021}
  [\href{https://arxiv.org/abs/2312.07612}{{\ttfamily 2312.07612}}].

\bibitem{Johns:2023xae}
L.~Johns and S.~Rodriguez, \emph{{Collisional flavor pendula and neutrino
  quantum thermodynamics}},  \href{https://arxiv.org/abs/2312.10340}{{\ttfamily
  2312.10340}}.

\bibitem{Johns:2019izj}
L.~Johns, H.~Nagakura, G.~M. Fuller and A.~Burrows, \emph{{Neutrino
  oscillations in supernovae: angular moments and fast instabilities}},
  \href{https://doi.org/10.1103/PhysRevD.101.043009}{\emph{Phys. Rev. D}
  {\bfseries 101} (2020) 043009}
  [\href{https://arxiv.org/abs/1910.05682}{{\ttfamily 1910.05682}}].

\bibitem{Padilla-Gay:2021haz}
I.~Padilla-Gay, I.~Tamborra and G.~G. Raffelt, \emph{{Neutrino Flavor Pendulum
  Reloaded: The Case of Fast Pairwise Conversion}},
  \href{https://doi.org/10.1103/PhysRevLett.128.121102}{\emph{Phys. Rev. Lett.}
  {\bfseries 128} (2022) 121102}
  [\href{https://arxiv.org/abs/2109.14627}{{\ttfamily 2109.14627}}].

\bibitem{Fiorillo:2023mze}
D.~F.~G. Fiorillo and G.~G. Raffelt, \emph{{Slow and fast collective neutrino
  oscillations: Invariants and reciprocity}},
  \href{https://doi.org/10.1103/PhysRevD.107.043024}{\emph{Phys. Rev. D}
  {\bfseries 107} (2023) 043024}
  [\href{https://arxiv.org/abs/2301.09650}{{\ttfamily 2301.09650}}].

\bibitem{Fiorillo:2023hlk}
D.~F.~G. Fiorillo and G.~G. Raffelt, \emph{{Flavor solitons in dense neutrino
  gases}}, \href{https://doi.org/10.1103/PhysRevD.107.123024}{\emph{Phys. Rev.
  D} {\bfseries 107} (2023) 123024}
  [\href{https://arxiv.org/abs/2303.12143}{{\ttfamily 2303.12143}}].

\bibitem{Wigner:1932eb}
E.~P. Wigner, \emph{{On the quantum correction for thermodynamic equilibrium}},
  \href{https://doi.org/10.1103/PhysRev.40.749}{\emph{Phys. Rev.} {\bfseries
  40} (1932) 749}.

\bibitem{Landau:1958joj}
L.~D. Landau, \emph{{On the Theory of the Fermi Liquid}},
  \href{https://doi.org/10.1016/b978-0-08-010586-4.50100-0}{\emph{J. Exp.
  Theor. Phys.} {\bfseries 35} (1958) }.

\bibitem{Johns:2023jjt}
L.~Johns, \emph{{Thermodynamics of oscillating neutrinos}},
  \href{https://arxiv.org/abs/2306.14982}{{\ttfamily 2306.14982}}.

\bibitem{Padilla-Gay:2020uxa}
I.~Padilla-Gay, S.~Shalgar and I.~Tamborra, \emph{{Multi-Dimensional Solution
  of Fast Neutrino Conversions in Binary Neutron Star Merger Remnants}},
  \href{https://doi.org/10.1088/1475-7516/2021/01/017}{\emph{JCAP} {\bfseries
  01} (2021) 017} [\href{https://arxiv.org/abs/2009.01843}{{\ttfamily
  2009.01843}}].

\bibitem{Zaizen:2023ihz}
M.~Zaizen and H.~Nagakura, \emph{{Characterizing quasisteady states of fast
  neutrino-flavor conversion by stability and conservation laws}},
  \href{https://doi.org/10.1103/PhysRevD.107.123021}{\emph{Phys. Rev. D}
  {\bfseries 107} (2023) 123021}
  [\href{https://arxiv.org/abs/2304.05044}{{\ttfamily 2304.05044}}].

\bibitem{Xiong:2023vcm}
Z.~Xiong, M.-R. Wu, S.~Abbar, S.~Bhattacharyya, M.~George and C.-Y. Lin,
  \emph{{Evaluating approximate asymptotic distributions for fast neutrino
  flavor conversions in a periodic 1D box}},
  \href{https://doi.org/10.1103/PhysRevD.108.063003}{\emph{Phys. Rev. D}
  {\bfseries 108} (2023) 063003}
  [\href{https://arxiv.org/abs/2307.11129}{{\ttfamily 2307.11129}}].

\bibitem{Abbar:2023ltx}
S.~Abbar, M.-R. Wu and Z.~Xiong, \emph{{Physics-informed neural networks for
  predicting the asymptotic outcome of fast neutrino flavor conversions}},
  \href{https://doi.org/10.1103/PhysRevD.109.043024}{\emph{Phys. Rev. D}
  {\bfseries 109} (2024) 043024}
  [\href{https://arxiv.org/abs/2311.15656}{{\ttfamily 2311.15656}}].

\bibitem{Cornelius:2023eop}
M.~Cornelius, S.~Shalgar and I.~Tamborra, \emph{{Perturbing fast neutrino
  flavor conversion}},
  \href{https://doi.org/10.1088/1475-7516/2024/02/038}{\emph{JCAP} {\bfseries
  02} (2024) 038} [\href{https://arxiv.org/abs/2312.03839}{{\ttfamily
  2312.03839}}].

\bibitem{Urquilla:2024bvf}
E.~Urquilla and S.~Richers, \emph{{Chaos in Inhomogeneous Neutrino Fast Flavor
  Instability}},  \href{https://arxiv.org/abs/2401.01936}{{\ttfamily
  2401.01936}}.

\end{thebibliography}\endgroup

\onecolumngrid
\appendix

\clearpage

\setcounter{equation}{0}
\setcounter{figure}{0}
\setcounter{table}{0}
\setcounter{page}{1}
\makeatletter
\renewcommand{\theequation}{S\arabic{equation}}
\renewcommand{\thefigure}{S\arabic{figure}}
\renewcommand{\thepage}{S\arabic{page}}

\begin{center}
\textbf{\large Supplemental Material for the Letter\\[0.5ex]
{\em Inhomogeneous Kinetic Equation for Mixed Neutrinos: Tracing the Missing Energy}}
\end{center}

In this Supplemental Material, we provide further details on the derivation of the equation of motion (EOM) discussed in the main text.

\bigskip


\section{A.~Detailed derivation of the equations of motion}

For the reader's convenience, we reproduce some of the equations from the main text. The starting point of the derivation is the many-body Hamiltonian describing the neutrino kinetic energy and interactions. This is written $\mathcal{H}=\mathcal{H}_0+\mathcal{U}$, where $\mathcal{H}_0=\sum_{\alpha,\bp}\epsilon_\bp a^\dagger_{\alpha,\bp}a_{\alpha,\bp}$ is the neutrino kinetic energy operator, and 
\begin{eqnarray}
    &&\mathcal{U}=\frac{\sqrt{2} \GF}{8}\sum_{\left\{\bp\right\},\alpha,\beta} a^\dagger_{\alpha,\bp_1}a_{\alpha,\bp_2}a^\dagger_{\beta,\bp_3}a_{\beta,\bp_4}\overline{u}_{\bp_1}\gamma^\mu u_{\bp_2}\overline{u}_{\bp_3}\gamma_\mu u_{\bp_4}.
\end{eqnarray}
We will henceforth use the notation $(\bp_1,\bp_2)(\bp_3,\bp_4)=\overline{u}_{\bp_1}\gamma^\mu u_{\bp_2}\overline{u}_{\bp_3}\gamma_\mu u_{\bp_4}$.

We want to determine the time evolution, under this Hamiltonian, of the density operator
\begin{equation}\label{eq:density_matrix}
    \mathcal{D}_{\alpha\beta}(\bp,\br)=\sum_\bk a^\dagger_{\beta,\bp+\bkk/2}a_{\alpha,\bp-\bkk/2}e^{-i\bk\cdot \br}, 
\end{equation}
or more conveniently of its spatial Fourier transform
\begin{equation}
    \tilde{\mathcal{D}}_{\alpha\beta}(\bp,\bk)=a^\dagger_{\beta,\bp+\bkk/2}a_{\alpha,\bp-\bkk/2}.
\end{equation}
Following the notation in the main text, we denote the expectation value over the quantum state $\langle \mathcal{D}_{\alpha\beta}(\bp,\br)\rangle=\rho_{\alpha\beta}(\bp,\br)$ and analogously for the Fourier transform \smash{$\langle \tilde{\mathcal{D}}_{\alpha\beta}(\bp,\bk)\rangle=\tilde{\rho}_{\alpha\beta}(\bp,\bk)$}.

We start by obtaining the commutator of the non-interacting Hamiltonian $\mathcal{H}_0$ with this operator; using the standard commutation rules we easily find
\begin{equation}
    [\mathcal{H}_0,\tilde{\mathcal{D}}_{\alpha\beta}(\bkk,\bp)]=\left(\epsilon_{\bp+\bkk/2}-\epsilon_{\bp-\bkk/2}\right) \tilde{\mathcal{D}}_{\alpha\beta}(\bkk,\bp).
\end{equation}
Expanding to first order in $\bkk$, and using ${\bm\nabla}_\bp\epsilon_\bp=\bv$,
we find 
\begin{equation}
    \left(\frac{\partial\tilde{\mathcal{D}}_{\alpha\beta}}{\partial t}\right)_0=i[\mathcal{H}_0,\tilde{\mathcal{D}}_{\alpha\beta}]=i\bv\cdot\bkk\,\tilde{\mathcal{D}}_{\alpha\beta},
\end{equation}
which Fourier transformed leads to the usual advective term quoted in the main text.

Let us now consider the commutator of the interaction Hamiltonian $\mathcal{U}$, which gives
\begin{eqnarray}
    [\mathcal{U},\tilde{\mathcal{D}}_{\alpha\beta}]&=&\frac{\sqrt{2}\GF}{8}\sum_{\left\{\bp\right\},\gamma}(\bp_1,\bp_2)(\bp_3,\bp_4)
    \nonumber\\ &&\kern5em
    {}\times\left[a^\dagger_{\gamma,\bp_1}a_{\gamma,\bp_2}a^\dagger_{\beta,\bp_3}a_{\alpha,\bp-\bkk/2}\delta_{\bp_4,\bp+\bkk/2}+a^\dagger_{\beta,\bp_1}a^\dagger_{\gamma,\bp_3}a_{\gamma,\bp_4}a_{\alpha,\bp-\bkk/2}\delta_{\bp_2,\bp+\bkk/2} \right. 
    \nonumber\\[1ex] && \kern5em
    \left.\kern1em{}-a^\dagger_{\beta,\bp+\bkk/2}a^\dagger_{\gamma,\bp_1}a_{\gamma,\bp_2}a_{\alpha,\bp_4}\delta_{\bp_3,\bp-\bkk/2}-a^\dagger_{\beta,\bp+\bkk/2}a_{\alpha,\bp_2}a^\dagger_{\gamma,\bp_3}a_{\gamma,\bp_4}\delta_{\bp_1,\bp-\bkk/2}
    \right],
\end{eqnarray}
where $\delta_{\bp_1,\bp_2}$ is a Dirac delta forcing the equality $\bp_1=\bp_2$. We can now take the expectation value of this operator over the mean-field state; the corresponding terms can be grouped in pairs which are identical after the exchange $\bp_1\to\bp_3$, $\bp_2\to \bp_4$, so we finally obtain
\begin{eqnarray}\label{eq:potential_commutator}
    \langle[\mathcal{U},\tilde{\mathcal{D}}_{\alpha\beta}]\rangle&=&\frac{\sqrt{2}\GF}{4}\sum_{\left\{\bp\right\},\gamma}(\bp_1,\bp_2)(\bp_3,\bp_4)
    \nonumber\\ &&\kern5em{}\times
    \left[(\langle a^\dagger_{\gamma,\bp_1}a_{\gamma,\bp_2}\rangle \langle a^\dagger_{\beta,\bp_3}a_{\alpha,\bp-\bkk/2}\rangle\right.
   -\langle a^\dagger_{\gamma,\bp_1}a_{\alpha,\bp-\bkk/2}\rangle \langle a^\dagger_{\beta,\bp_3}a_{\gamma,\bp_2}\rangle)\delta_{\bp_4,\bp+\bkk/2}
   \nonumber\\[1ex]&& \kern5em
   \left.\kern1em{}-(\langle a^\dagger_{\gamma,\bp_1}a_{\gamma,\bp_2}\rangle \langle a^\dagger_{\beta,\bp+\bkk/2}a_{\alpha,\bp_4}\rangle-\langle a^\dagger_{\gamma,\bp_1}a_{\alpha,\bp_4}\rangle \langle a^\dagger_{\beta,\bp+\bkk/2}a_{\beta,\bp_4}\rangle)\delta_{\bp_3,\bp-\bkk/2}  \right].
\end{eqnarray}

In each term, it will prove convenient to reparametrize the momenta in such a way as to separate the large momenta (associated with the neutrino energies) from the small momenta (associated with the spatial gradients of the density matrix). In particular, in terms of the form $\langle a^\dagger_{\beta,\bp_1}a_{\alpha,\bp_2}\rangle$, because of the slowly-varying nature of the density matrix, we find a non-vanishing result only if $\bp_1$ and $\bp_2$ differ by terms of order $\sim \ell^{-1}$. Thus, for example, it proves convenient to reparameterize the first term $\langle a^\dagger_{\gamma,\bp_1}a_{\gamma,\bp_2}\rangle \langle a^\dagger_{\alpha,\bp_3}a_{\beta,\bp-\bkk/2}\rangle$ choosing $\bp_1=\bp'+(\bkk-\bq)/4$, $\bp_2=\bp'-(\bkk-\bq)/4$, $\bp_3=\bp+\bq/2$, $\bp_4=\bp+\bkk/2$, which makes it clear that $\bp$ and $\bp'$ are large momenta, while $\bq$ and $\bkk$ are small. Adopting a similar parameterization for all the terms we find
\begin{eqnarray}
    \langle[\mathcal{U},\tilde{\mathcal{D}}_{\alpha\beta}]\rangle&=&\frac{\sqrt{2}\GF}{4}\sum_{\bp',\bq,\gamma} \left[\tilde{\rho}_{\alpha\beta}\left(\bp+\frac{\bq-\bkk}{4},\frac{\bq+\bkk}{2}\right)\tilde{\rho}_{\gamma\gamma}\left(\bp',\frac{\bk-\bq}{2}\right)\right.\left(\bp+\frac{\bq}{2},\bp+\frac{\bkk}{2}\right)\left(\bp'+\frac{\bkk-\bq}{4},\bp'+\frac{\bq-\bkk}{4}\right)
    \nonumber\\[1ex] &&\kern2em
    {}-\tilde{\rho}_{\alpha\gamma}\left(\bp+\frac{\bq-\bkk}{4},\frac{\bq+\bkk}{2}\right)\tilde{\rho}_{\gamma\beta}\left(\bp',\frac{\bkk-\bq}{2}\right) \left(\bp+\frac{\bq}{2},\bp'+\frac{\bq-\bkk}{4}\right)\left(\bp'+\frac{\bkk-\bq}{4},\bp+\frac{\bk}{2}\right)
    \nonumber\\[1ex] && \kern2em
    {}-\tilde{\rho}_{\alpha\beta}\left(\bp+\frac{\bkk-\bq}{4},\frac{\bkk+\bq}{2}\right)\tilde{\rho}_{\gamma\gamma}\left(\bp',\frac{\bkk-\bq}{2}\right) \left(\bp-\frac{\bkk}{2},\bp-\frac{\bq}{2}\right)\left(\bp'+\frac{\bkk-\bq}{4},\bp'+\frac{\bq-\bkk}{4}\right)
    \nonumber\\[1ex]  &&\kern2em
    {}+\tilde{\rho}_{\alpha\gamma}\left(\bp',\frac{\bkk-\bq}{2}\right)\tilde{\rho}_{\gamma\beta}\left(\bp+\frac{\bkk-\bq}{4},\frac{\bkk+\bq}{2}\right) \left.\left(\bp-\frac{\bkk}{2},\bp'+\frac{\bq-\bkk}{4}\right)\left(\bp'+\frac{\bkk-\bq}{4},\bp-\frac{\bq}{2}\right)\right].
\end{eqnarray}

We can now take the inverse Fourier transform by multiplying by $e^{-i\bkk\cdot\br_1}$ and summing over all $\bkk$. We perform the full manipulations only on the first of the four terms; the other ones are treated analogously. Introducing the notation $\ba=\bkk-\bq$, $\bb=\bkk+\bq$, we can write this term as
\begin{eqnarray}\nonumber
    &&\sum_{\frac{\ba}{2},\frac{\bb}{2},\bp',\gamma}e^{i\frac{\bb}{2}\cdot(\br_2-\br_1)+i\frac{\ba}{2}\cdot(\br_3-\br_1)}\rho_{\gamma\gamma}(\bp',\br_3)\rho_{\alpha\beta}\left(\bp-\frac{\ba}{4},\br_2\right)\left(\bp+\frac{\bq}{2},\bp+\frac{\bk}{2}\right)\left(\bp'+\frac{\ba}{4},\bp'-\frac{\ba}{4}\right).
\end{eqnarray}
In this expression, we must now perform the expansion for $|\bk,\bq|\ll|\bp,\bp'|$.
For the spinor part, this requires using the explicit form of the spinors expanded to first order in the momenta $\ba$ and $\bb$. The result depends on the mutual orientation of the vectors $\bp$ and $\bp'$. Thus, we introduce a set of three orthonormal directions
\begin{equation}
    \be_y=\frac{\bp\times\bp'}{p p' \sin \theta_{\bp\bp'}},\quad
    \be_x=\frac{\be_y\times\bp}{p},\quad
    \be_z=\frac{\bp}{p}.
\end{equation}
We can now use the result that
\begin{eqnarray}
    &&\left(\bp+\frac{\bq}{2},\bp+\frac{\bk}{2}\right)\left(\bp'+\frac{\ba}{4},\bp'-\frac{\ba}{4}\right)\simeq 4(1-\cos\theta_{\bp\bp'}) -\frac{\sin\theta_{\bp\bp'}[b_x p'+i a_y (p+p')]}{p p'}.
\end{eqnarray}
This expression is obtained by a direct use of the corresponding spinors.
The other matrix element is
\begin{equation}
    \left(\bp+\frac{\bq}{2},\bp'-\frac{\ba}{4}\right)\left(\bp'+\frac{\ba}{4},\bp+\frac{\bk}{2}\right)=-\left(\bp+\frac{\bq}{2},\bp+\frac{\bk}{2}\right)\left(\bp'+\frac{\ba}{4},\bp'-\frac{\ba}{4}\right)
\end{equation}
by virtue of Fierz identities. 

The expansion in powers of $\bk$ and $\bq$ involves three terms. The zeroth-order term is then simply
\begin{equation}
    4\sum_{\bp',\gamma}(1-\cos\theta_{\bp\bp'})\rho_{\gamma\gamma}(\bp',\br)\rho_{\alpha\beta}(\bp,\br),
\end{equation}
where we call $\br_1=\br$ since only one spatial variable survives in the final result. Next we have a first-order term coming from the gradient expansion $\rho_{\alpha\beta}(\bp-\ba/4,\br_2)\simeq \rho_{\alpha\beta}(\bp,\br_2)-{\bm\nabla}_\bp \rho_{\alpha\beta}\cdot\ba/4$, leading to
\begin{equation}\label{eq:term_derivatives}
    -2i\sum_{\bp',\gamma}(1-\cos\theta_{\bp\bp'}){\bm\nabla}_\br \rho_{\gamma\gamma}(\bp',\br) \cdot {\bm\nabla}_\bp \rho_{\alpha\beta}(\bp,\br).
\end{equation}
Finally, from the expansion of the spinor part we obtain
\begin{equation}
    -\sum_{\bp',\gamma}\frac{\sin\theta_{\bp\bp'}}{pp'}\Bigl[2ip'\rho_{\gamma\gamma}(\bp',\br)\,\be_x\cdot{\bm\nabla}_\br \rho_{\alpha\beta}(\bp,\br)-2(p+p') \rho_{\alpha\beta}(\bp,\br)\,
    \be_y\cdot{\bm\nabla}_\br \rho_{\gamma\gamma}(\bp',\br)\Bigr].
\end{equation}
\exclude{
\begin{eqnarray}
    -&&\sum_{\bp',\gamma}\frac{\sin\theta_{\bp\bp'}}{pp'}\left[2ip'\rho_{\gamma\gamma}(\bp',\br)\be_x\cdot\partial_\bx \rho_{\alpha\beta}(\bp,\br)\right.\\ \nonumber &&\left.-2(p+p')\be_y\cdot\partial_\bx \rho_{\gamma\gamma}(\bp',\br)\rho_{\alpha\beta}(\bp,\br)\right].
\end{eqnarray}
}
Performing analogous manipulations on all four terms of Eq.~\eqref{eq:potential_commutator} we finally reach the form
\begin{equation}
    \frac{\partial \rho_{\bp,\br}}{\partial t}+\bv\cdot{\bm\nabla}_\br\rho_{\bp,\br}=i\sqrt{2}\GF\sum_{\bp'}(1-\cos\theta_{\bp\bp'})[\rho_{\bp,\br},\rho_{{\bp',\br}}]+\left(\frac{\partial \rho_{\bp,\br}}{\partial t}\right)_\mathrm{der}+\left(\frac{\partial \rho_{\bp,\br}}{\partial t}\right)_\mathrm{spinor},
\end{equation}
where we introduce the compact notation $\rho_{\bp,\br}=\rho(\bp,\br)$ as in the main text. The terms $(\partial\rho_{\bp,\br}/\partial t)_\mathrm{der}$ originate from the derivatives of the density matrices with respect to momenta, analogously to Eq.~\eqref{eq:term_derivatives}, and can be written as
\begin{equation}
    \left(\frac{\partial \rho_{\bp,\br}}{\partial t}\right)_\mathrm{der}=\frac{\sqrt{2}\GF}{2}\sum_{\bp'}(1-\cos\theta_{\bp\bp'})\left(\Bigl\{{\bm\nabla}_\br\rho_{\bp',\br},
    {\bm\nabla}_\bp \rho_{\bp,\br}\Bigr\}+2\mathrm{Tr}({\bm\nabla}_\br\rho_{\bp',\br})\cdot
    {\bm\nabla}_\bp \rho_{\bp,\br}\right).
\end{equation}
The expansion of the spinors leads to a term of the form
\begin{eqnarray}
    \left(\frac{\partial \rho_{\bp,\br}}{\partial t}\right)_\mathrm{spinor}&=&
    \frac{\sqrt{2}\GF}{2}\sum_{\bp'}\frac{\sin\theta_{\bp\bp'}\be_x}{p}\cdot
    \left(\Bigl\{\rho_{\bp',\br},{\bm\nabla}_\br\rho_{\bp,\br}\Bigr\}
    +2\mathrm{Tr}(\rho_{\bp',\br}){\bm\nabla}_\br\rho_{\bp,\br}\right)
    \nonumber\\ &&{}+\frac{i\sqrt{2}\GF}{2}\sum_{\bp'}\sin\theta_{\bp\bp'}\left(\frac{1}{p}+\frac{1}{p'}\right)\be_y\cdot
    \Bigl[\rho_{\bp,\br},{\bm\nabla}_\br\rho_{\bp',\br}\Bigr].
\end{eqnarray}
Using now the identity
\begin{equation}
    {\bm\nabla}_\bp\cos\theta_{\bp\bp'}=\frac{\sin\theta_{\bp\bp'}}{p}\,\be_x,
\end{equation}
and repeatedly integrating by parts, we are led to the equations finally reported in the main text.

\section{B.~Entropy Conservation}

Besides conserving energy, the newly introduced EOMs have also the property of conserving the entropy of the neutrino gas, defined by
\begin{equation}
    S_\nu=\int d^3\br \sum_{\bp}\mathrm{Tr}\left(s_{\bp,\br}\right),
\end{equation}
where
\begin{equation}
    s_{\bp,\br}=-\Bigl[\rho_{\bp,\br}\log \rho_{\bp,\br}+(\mathbb{1}-\rho_{\bp,\br})\log(\mathbb{1}-\rho_{\bp,\br})\Bigr],
\end{equation}
so that the time derivative can be written 
\begin{equation}\label{eq:entropy_change}
    \frac{dS_\nu}{dt}=\int d^3\br \sum_{\bp}\mathrm{Tr}\left[\frac{\partial\rho_{\bp,\br}}{\partial t}\log\left(\frac{\mathbb{1}-\rho_{\bp,\br}}{\rho_{\bp,\br}}\right)\right].
\end{equation}
For the standard EOMs, without the new terms, entropy conservation is automatically enforced. We now show that the new terms from the gradient expansion also maintain this conservation law.

To compute the time derivative of the entropy, it is convenient to rewrite the EOMs reported in the main text as
\begin{equation} \label{eq:full_eom_schematic}
    \partial_t \rho_{\bp,\br}+\bv\cdot{\bm\nabla}_\br \rho_{\bp,\br}
    =i\Bigl[\rho_{\bp,\br},{\sf\Omega}_{\bp,\br}\Bigr]
    +\frac{1}{2}\left\{{\bm\nabla}_\br {\sf\Omega}^{(0)}_{\bp,\br},{\bm\nabla}_\bp \rho_{\bp,\br}\right\} 
    -\frac{1}{2}\left\{{\bm\nabla}_\bp {\sf\Omega}^{(0)}_{\bp,\br},{\bm\nabla}_\br \rho_{\bp,\br}\right\}.
\end{equation}
We can now replace the time derivative Eq.~\eqref{eq:full_eom_schematic} in the rate of change of the entropy Eq.~\eqref{eq:entropy_change}; the advection term of course does not lead to any change in entropy. The commutator term also leads to a vanishing entropy change, as follows from the identity
$\mathrm{Tr}\left([{\sf\Omega}_{\bp,\br},\rho_{\bp,\br}]\,{\sf F}(\rho_{\bp,\br})\right)=0$, where the matrix ${\sf F}$ is a generic function. Finally, from the anticommutator terms we find
\begin{equation}
    \frac{dS_\nu}{dt}=\frac{1}{2}\int d^3\br\sum_{\bp} \mathrm{Tr}\left[\left\{{\bm\nabla}_\br {\sf\Omega}^{(0)}_{\bp,\br},{\bm\nabla}_\bp s_{\bp,\br}\right\}-\left\{{\bm\nabla}_\bp {\sf\Omega}^{(0)}_{\bp,\br},{\bm\nabla}_\br s_{\bp,\br}\right\}\right],
\end{equation}
where we have recognized the total derivative
\begin{equation}
    {\bm\nabla}_\bp \rho_{\bp,\br}\log\left(\frac{\mathbb{1}-\rho_{\bp,\br}}{\rho_{\bp,\br}}\right)={\bm\nabla}_\bp s_{\bp,\br}.
\end{equation}
Even though the operators $\rho_{\bp,\br}$ and $h_{\bp,\br}$ do not commute, this manipulation can be performed for all terms within the trace. Finally, by integrating by parts both terms, we can easily conclude that the rate of change of the entropy vanishes. Therefore, we recover the physically intuitive conclusion that, in the absence of incoherent collisions, which could only appear to second order in $\GF$, neutrino entropy is conserved.

\end{document}